\documentclass[12pt]{iopart}

\usepackage{iopams}  

\usepackage{graphicx}
\usepackage{dcolumn}
\usepackage{bm}

\usepackage{xcolor}

\begin{document}

\title{Emulation of the final \textit{r}-process abundance pattern with a neural network}

\author{Yukiya Saito\footnote{Current address: Department of Physics, University of Notre Dame, Notre Dame, IN 46556, USA}}
 \ead{yukiya@alum.ubc.ca}
\address{%
 TRIUMF, 4004 Wesbrook Mall, Vancouver, BC V6T 2A3, Canada
}%
\address{%
 Department of Physics and Astronomy, The University of British Columbia, Vancouver, BC V6T 1Z1, Canada
}%

\author{Iris Dillmann}
\address{TRIUMF, 4004 Wesbrook Mall, Vancouver, BC V6T 2A3, Canada}
\address{Department of Physics and Astronomy, University of Victoria, Victoria, BC V8P 5C2, Canada}

\author{Reiner Kr{\"u}cken}
\address{TRIUMF, 4004 Wesbrook Mall, Vancouver, BC V6T 2A3, Canada}
\address{Department of Physics and Astronomy, The University of British Columbia, Vancouver, BC V6T 1Z1, Canada}
\address{Nuclear Science Division, Lawrence Berkeley National Laboratory,
Berkeley, California 94720, USA}

\author{Matthew R. Mumpower}
\address{Theoretical Division, Los Alamos National Laboratory, Los Alamos, NM 87545, USA}
\address{Center for Theoretical Astrophysics, Los Alamos National Laboratory, Los Alamos, NM 87545, USA}

\author{Rebecca Surman}
\address{Department of Physics, University of Notre Dame, Notre Dame, IN 46556, USA}

\date{\today}

\begin{abstract}
This work explores the construction of a fast emulator for the calculation of the final pattern of nucleosynthesis in the rapid neutron capture process (the $r$-process). An emulator is built using a feed-forward artificial neural network (ANN). We train the ANN with nuclear data and relative abundance patterns. We take as input the $\beta$-decay half-lives and the one-neutron separation energy of the nuclei in the rare-earth region. The output is the final isotopic abundance pattern. In this work, we focus on the nuclear data and abundance patterns in the rare-earth region to reduce the dimension of the input and output space.
We show that the ANN can capture the effect of the changes in the nuclear physics inputs on the final $r$-process abundance pattern in the adopted astrophysical conditions. We employ the deep ensemble method to quantify the prediction uncertainty of the neutal network emulator. The emulator achieves a speed-up by a factor of about 20,000 in obtaining a final abundance pattern in the rare-earth region. The emulator may be utilized in statistical analyses such as uncertainty quantification, inverse problems, and sensitivity analysis.
\end{abstract}



\maketitle

\section{\label{sec:Intro}Introduction}
In the studies of heavy element nucleosynthesis, especially the rapid neutron capture process ($r$-process), it is widely recognized that the properties of atomic nuclei, e.g. masses, shell structures, decay half-lives, and $\beta$-delayed neutron emission probabilities, affect the prediction of the resulting abundance pattern \cite{B2FH,AlCameron,CowanReview}.

A common method to investigate the impact of nuclear physics inputs on the $r$-process abundance pattern is to run a large number of nuclear reaction network calculations while varying the relevant inputs, e.g. $\beta$-decay half-lives ($T_{1/2}$), neutron separation energies ($S_n$), neutron capture rates, etc., following the design of the numerical experiment \cite{Mumpower2016, Martin2016, Sprouse2020, Rauscher2016, Nishimura2017,Bliss2020}.
The resulting calculated abundance patterns can then be used to obtain Monte Carlo estimates of the propagated uncertainty and sensitivity of the varied inputs.
While the nuclear reaction network calculations can be run in parallel (in a so-called \emph{embarrassingly parallel} scheme) for these purposes, the computational cost will still be significant.
If one hopes to include a large number of inputs in the Monte Carlo studies, the required number of data points exponentially grows due to ``the curse of dimensionality''. In variance-based sensitivity analyses (e.g., Ref.~\cite{Kiss_BRIKEN_2022}), the problem is even more challenging since a ``sufficient'' number of unique data points is required for each of the input variables of interest, and it is difficult, if not impossible, to know \emph{a priori} how many data points are sufficient.

Another class of statistical analysis using nuclear reaction network calculations is solving inverse problems, that is, to find the optimal values of the nuclear physics inputs that best reproduce the observed solar $r$-process abundance pattern \cite{Mumpower_reverse_2017,Vassh_reverse_2021}. 
In these works, the Markov chain Monte Carlo (MCMC) method was used to minimize the $\chi^2$ likelihood between the calculated and observed solar abundance patterns in the rare-earth region, by adjusting the correction to the theoretical nuclear masses.
While MCMC is a powerful tool for solving inverse problems, the computational cost can become prohibitively large. This is because a large number of samples have to be generated by solving forward problems (in the current context, by performing nuclear reaction network calculations) to explore the entire parameter space. The generation of samples cannot be easily parallelized as a step in MCMC depends on the previous step. Since a nuclear reaction network calculation typically takes at least a few minutes, it could take an extremely long time before sufficient statistics are obtained. 
To mitigate this problem, Refs.~\cite{Mumpower_reverse_2017,Vassh_reverse_2021} reduced the number of independent input variables by parameterizing the correction to the mass surface to reproduce the detailed features of the rare earth peak. 

Another way to tackle the problem is to reduce the computational cost.
This can be achieved by creating an emulator, that is, modeling the response of the output to the variation of the inputs. In this way, an expensive computer code is replaced by a fast emulator. In nuclear physics, some of the common approaches for building an emulator are to replace the original computer models with reduced-order models, Gaussian processes (GPs), or neural networks \cite{Frame2018,Drischler2021,Drischler2022,Bonilla2022,Melendez2022,Garcia2023,Giuliani2023,Odell2024,Lay2024}. 

In this work, our aim is to model the map between nuclear-physics inputs to the nuclear reaction network calculations and their output, i.e. calculated abundances. Since both input and output are high-dimensional, we employ artificial neural networks which can work well with large input and output dimensions \cite{Goodfellow_DL_2016}. We also introduce a way to quantify the uncertainty associated with the emulator using a technique called deep ensembles~\cite{DeepEnsemble}.

The organization of this paper is as follows. In Section~\ref{sec:Theory}, the relevant theoretical background is introduced. In Section~\ref{sec:NumericalExperiment}, we describe how the artificial neural network emulator is constructed for nuclear reaction network calculations, and the results are reported in Section~\ref{sec:Results}. In Section~\ref{sec:example}, we use the emulator to propagate the uncertainty of the nuclear physics input to the nucleosynthesis yields. Finally, conclusions are given in Section~\ref{sec:Conclusions}.

\section{\label{sec:Theory}Theoretical background}

\subsection{\label{subsec:ReactionNetwork}Nuclear Reaction Network Calculation}
Nuclear reaction networks are often used to study the evolution of nuclear abundances during an astrophysical event \cite{Lippuner_SkyNet_2017, Reichert2023, SprouseTheis2020}. 
From the evolution of nuclear abundances, it is also possible to compute the amount of energy released from nuclear decays and reactions. The calculation of energy release is important for modeling the reheating of the material and the kilonova lightcurves \cite{Zhu2018,Zhu2021,Barnes2021,Lund2023,Holmbeck2023}. 

Detailed nucleosynthesis calculations require information on the temperature and/or density of the astrophysical medium. This information can be obtained from tracer particles in hydrodynamical simulations of astrophysical events of interest. Tracer particles record the evolution of the positions and thermodynamic properties of the astrophysical plasma. The temporal evolution of temperature and/or density of each tracer particle is often referred to as an astrophysical trajectory. The composition of nuclear species at the beginning of the astrophysical trajectory may be obtained from the hydrodynamical simulation itself or by assuming nuclear statistical equilibrium (NSE) at a sufficiently high temperature. The initial composition is then evolved through a nuclear reaction network calculation along the astrophysical trajectory.


To obtain the evolution of nuclear abundances, astrophysical trajectories are divided into small time steps and the following system of ordinary differential equations is solved (integrated) for nuclear abundances at each time step:
\begin{eqnarray}
    \frac{\mathrm{d}Y_i}{\mathrm{d}t} &= \sum_{j} \mathcal{N}^i_j \lambda_j Y_j + \sum_{j,k} \mathcal{N}^i_{j,k}\; \rho N_A \langle{j,k}\rangle Y_j Y_k \nonumber \\
    &+\sum_{j,k,l} \mathcal{N}^i_{j,k,l}\; \rho^2 {N_A}^2 \langle{j,k,l}\rangle Y_j Y_k Y_l, \label{eq:ReactionNetwork}
\end{eqnarray}
where $Y_i$ is the nuclear abundance of the nucleus $i$, defined so that $\sum_{i} Y_i A_i = \sum_{i} X_i = 1$, with $X_i$ being the mass fraction of the nucleus $i$ and $A_i$ the respective mass number, and $\rho$ is the density of the astrophysical medium, which is obtained from hydrodynamical simulations. The factors $\mathcal{N}^i_{j}$, $\mathcal{N}^i_{j,k}$, and $\mathcal{N}^i_{j,k,l}$ account for how many particles of nucleus $i$ are created or destroyed in the reaction, while correcting for over-counting due to having the same nuclear species in a reaction. $\lambda_i$, $N_A \langle{j,k}\rangle$, and $N_A \langle{j,k,l}\rangle$ are the one-body reaction or decay rate, two-body reaction rate between nuclei $i$ and $j$, and three-body reaction rate between nuclei $j$, $k$, and $l$. One-body reactions include nuclear decays and reactions of nuclei with photons (photodissociation), electrons (electron capture), and neutrinos. The one-body, two-body, and three-body reaction rates typically depend on the temperature of the medium, which can be given by hydrodynamical simulations, but it is also possible to compute the temperature at each time step using the density and entropy of the system, assuming adiabatic or radiation dominated gas. 

Performing a nuclear reaction network calculation and obtaining the final abundance pattern according to the system of coupled ordinary differential equations (ODEs) is an initial value problem with respect to the initial conditions. This system of ODEs is considered \emph{stiff}, since the decay and reaction rates vary drastically and there the changes in abundance in each time step ranges over many orders of magnitude. For a stiff system of ODE, the \emph{implicit Euler} method is often used to obtain the evolution of the abundances. For details of nuclear reaction network calculations, see e.g. Refs.~\cite{Lippuner_SkyNet_2017, Reichert2023, SprouseTheis2020, Hix_network_2006}.

In this work, we use an implementation of nuclear reaction network called \textsc{PRISM} \cite{SprouseTheis2020}. Although there exist several numerical softwares for abundance calculations, \textsc{PRISM} offers a straightforward way to manage and manipulate the nuclear physics data necessary for the calculation, which is ideal for this application. The baseline nuclear data are identical to the ones used in Ref.~\cite{Kiss_BRIKEN_2022}, which mainly consists of nuclear physics models developed at Los Alamos National Laboratory (LANL) \cite{FRDM2012, Moller_qrpa_2019, Kawano_COH_2021}. 

As a result of a nuclear reaction network calculation, the temporal evolution of nuclear abundances can be obtained. In the case of the $r$-process, thousands of nuclear species -- ranging from neutron to actinides and possibly beyond, are involved. It is an extremely complex task to emulate the full time dependence of nuclear abundances across the chart of nuclides. Therefore, in this work, we limit our focus to the nuclei in the rare-earth region, whose mass numbers range from $A=150$ to $A=180$. We also focus on emulating their localized effect on the final abundance patterns. The observed abundance pattern of the rare-earth nuclei is known as the rare-earth peak, and the details are discussed in the following section. Emulation of the temporal evolution of abundances will require further development and will be a future work.

\subsection{\label{subsec:RareEarthPeak}The Rare Earth Peak}


In the late time of the $r$-process, when the temperature and the available number of neutrons decrease, the material starts to decay towards stability. This is called the $r$-process \emph{freeze-out}. During the freeze-out, the timescales of neutron captures and $\beta$-decays become similar, and the balance of these processes may determine some of the features of the abundance pattern, including the \emph{rare-earth peak} (REP, $A\sim$165) \cite{Surman_rep_1997,Mumpower_rep_2012a,Mumpower_rep_2012b,Mumpower_rep_2012c}. Fission during the freeze-out may also have a significant impact on the formation of the rare-earth peak, as well as the second ($A\sim130$) and third ($A\sim195$) abundance peaks \cite{Panov_repfission_2008,Goriely_repfission_2013,Eichler_repfission_2015}. Furthermore, some neutron-rich nuclei undergo (one or multiple) neutron emissions following $\beta$-decays ($\beta$-delayed neutron emission). Since this not only alters the path of the decay chains towards stability, but also provides free neutrons during the freeze-out, it can have a significant effect on the final abundance pattern.

According to this picture, understanding the synthesis of the lanthanides ($A=150$-180) in this mass region may allow us to probe the detailed conditions of the freeze-out and the mechanisms of the $r$-process that robustly reproduce the abundance pattern occurring in stars over a wide range of metallicities \cite{Surman_rep_1997, Mumpower_rep_2012a}. 

The formation of the REP is sensitive to variables that control the neutron density and neutron-to-seed ratio in the late stages of the $r$-process, such as the timescale for the expansion of the material. However, these astrophysical conditions are entangled with nuclear physics processes that provide additional neutrons, of which $\beta$-delayed neutron emissions can be a main contributor \cite{Arcones2011a}. The mass region and nuclei responsible for the formation of the REP have previously been inferred \cite{Mumpower_rep_2012a}. However, the most important nuclei lie about 10--15 mass units away from the valley of stability, and the experimental knowledge of $\beta$-decay properties for these neutron-rich isotopes has so far been very limited \cite{Kiss_BRIKEN_2022}.   

Several authors have proposed that during the $r$-process freeze-out the competition between $\beta^-$-decays and neutron captures shape the REP while the material decays back to stability \cite{Surman_rep_1997, Mumpower_rep_2012a,Mumpower_rep_2012b,Mumpower_rep_2012c, Arcones2011a, Surman2001}. Neutron emission following $\beta^-$-decays of neutron-rich nuclei may also have a significant impact on the abundance pattern by providing additional neutrons to the environment and changing the mass number of the nuclide. 
Nuclear masses are also relevant in the form of reaction/decay Q-values. Especially, one neutron separation energy $S_{n}$ directly affects the photodissociation rates, which are calculated from neutron capture rates via detailed balance:
\begin{eqnarray}
    \lambda_{(\gamma,n)} = &\left\langle{\sigma v}_{(n,\gamma)}\right\rangle \cdot \frac{G(N,Z)\cdot G(1,0)}{G(N+1,Z)} \cdot \left( \frac{A}{A+1} \right)^{3/2} \nonumber \\
    &\cdot \left( \frac{m_u kT}{2\pi\hbar^2} \right)^{3/2} \cdot \exp \left( -\frac{S_{n}(N+1,Z)}{kT} \right), \label{eq:detailedbalance}
\end{eqnarray}
where $\left\langle{\sigma v}_{(n,\gamma)}\right\rangle$ is the velocity-integrated neutron capture cross section for a nucleus with $N$ neutrons and $Z$ protons ($A\equiv N+Z$), $G(N,Z)$ is the partition function for the nucleus $(N,Z)$ ($G(1,0)$ is the partition function for neutron), $m_u$ is the mass of a nucleon, and $T$ is the temperature of the environment. The detailed balance implies that a change in $S_n$ affects the distribution of abundances in each isotopic chain during $(n, \gamma) \leftrightarrows (\gamma, n)$ equilibrium. Change in the photodissociation rates may also affect the net flow of neutron capture at the onset of freeze-out when the temperature is still sufficiently high for the photodissociation to be active.

In this work, we focus on the effect of the $\beta^-$-decay rates and one-neutron separation energies on the REP. The effect of $\beta$-delayed neutron emission is indirectly taken into account through the variation of $\beta^-$-decay rates.


\subsection{\label{subsec:NeuralNetwork}Feed-forward Artificial Neural Network}
A feed-forward artificial neural network (ANN) can be described as a series of functional transformations, where an input (vector) $\bm{x}$ is propagated through intermediate layers and finally to the output (vector) $\bm{y}$ \cite{Goodfellow_DL_2016, Bishop_pattern_2006}. 
For example, if an ANN has three layers that are connected in a chain, it can be expressed as $f(\bm{x}) = f^{(3)} \left( f^{(2)}\left(f^{(1)}(\bm{x}) \right) \right)$.
An ANN with trainable parameters, or \emph{weights} $\bm{w}$, defines a mapping $\bm{y}=f\left(\bm{x}, \bm{w}\right)$, which is trained to approximate some function $\bm{y} = f^* (\bm{x})$. Training is typically performed using variations of gradient descent algorithms to minimize cost functions, such as the mean squared error cost function. The information from the cost function is propagated backwards through the network to compute the gradient of the cost function with respect to the trainable parameters, using the back-propagation algorithm. Much of the design of architectures of ANNs goes into the choice of type, number (depth) and width of layers, and how each layer is connected. In what follows, the two types of layers used in the design of the current architecture of ANNs are described. Please see Ref.~\cite{Goodfellow_DL_2016} for more details.

We select the ANN library called \textsc{Keras} \cite{Keras} as the implementation of the neural network, which provides a high level application programming interface (API) for \textsc{TensorFlow} \cite{Tensorflow} in \textsc{Python}. 

\subsubsection{Fully Connected Layers}
In a fully connected (or dense) layer, linear combinations of the input of the layer are first constructed. Taking the first layer of the ANN as an example and following the notation in \cite{Bishop_pattern_2006}, the linear combination can be written as
\begin{equation}
    a_j = \sum_{i=1}^D w_{ji}^{(1)} x_i + w_{j0}^{(1)},
\end{equation}
where $j=1,\ldots, M$ with $M$ being the dimension of the layer, $D$ the dimension of the input $\bm{x} = x_1, \ldots, x_D$, $w_{ji}^{(1)}$ the weights with the superscript (1) denoting the first layer, and $w_{j0}^{(1)}$ the biases. Each of $a_j$ is subsequently transformed using nonlinear activation function $h(\cdot)$ to obtain outputs of the layer
\begin{equation}
    z_j = h(a_j). 
\end{equation}
The activation function is chosen to be the rectified linear unit (ReLU) $h(a_j) = \max{\{0, a_j\}}$ \cite{ReLU}, which is one of the most widely used activation functions. The subsequent layers take the output of the previous layers, and the functional transformations can be operated in the same manner. In the final layer, the identity activation function (where the output of the function is identical to the input) is used to allow for unbounded values. 

\subsubsection{Convolutional Layers}
Convolutional neural networks (CNNs) are a type of ANNs where convolution is used in at least one of the layers. CNNs are used most extensively in the field of computer vision. The central assumption in CNNs is that nearby pixels in the image data or neighboring data points in time-series data are strongly correlated \cite{Goodfellow_DL_2016,Bishop_pattern_2006}. Based on this assumption, it is possible to extract the local features in the data. The general convolution operation for multidimensional arrays can be expressed as \cite{Goodfellow_DL_2016}
\begin{eqnarray}
    S(i,j) &= (I*K) (i,j) \nonumber \\
    &= \sum_{m}\sum_{n} I(m,n) K(i-m, j-n),
\end{eqnarray}
where $I$ is a multidimensional array image input, $K$ is called a kernel (which is a multidimensional array that stores the adaptive weights), and $S(i,j)$ is the output and is often referred to as a feature map. 

In the actual implementation of convolution for machine learning purposes, convolution is typically performed for a small subregion in an input image for each element in a feature map. Furthermore, the weights in a feature map are used for all the elements in an input, which is the concept called ``weight (parameter) sharing''. These reduce the number of weights that are stored in the memory, making it possible to process images that have a large number of pixels. Each unique set of weights is often referred to as a filter, which acts to detect different features in the input data.

The use of convolutional layers in this work is motivated by the expectation that the properties of neighboring nuclei on the chart of nuclides have a correlated effect on the final abundance pattern. As discussed in more detail in Section~\ref{subsec:Arch}, superior performance was obtained when convolutional layers were used.

\subsubsection{Uncertainty quantification with deep ensembles}
In emulation of computer codes, one of the most popular approaches is to use a probabilistic model called Gaussian processes (GPs) \cite{GP, Conti_GPEmulator_2009,Bastos_GPEmulator_2009,Kennedy_GPEmulator_2006}. GPs provide a natural way to quantify the prediction uncertainty in terms of the quality of the emulation. On the other hand, typical ANNs including CNNs have deterministic weights, therefore, their predictions are also deterministic. An approach to overcome this limitation for ANNs is to employ Bayesian neural networks (BNNs) \cite{Neal_BNN_2012} where the weights are expressed as probability distributions. Similarly to GPs, it is possible to quantify the prediction uncertainty of BNNs due to their probabilistic nature. However, the computational complexity of BNNs is large compared to traditional ANNs, and the application to CNNs is technically challenging for non-experts. 

The approach we employ in this work to quantify prediction uncertainty is called \emph{deep ensemble}, which is a simple and scalable method and has been shown to provide robust and accurate estimates of uncertainty, comparable to BNNs \cite{DeepEnsemble}. In a deep ensemble, $M$ (ensemble size, here, $M=10$) copies of the same ANN architecture are used and each of them is randomly initialized. The final layer of the original ANN architecture is replaced with a layer with two outputs: a predictive mean $\mu(\bm{x})$ and a predictive variance $\sigma^2(\bm{x})$.
The predictive mean and variance of the ensemble are obtained by treating the ensemble of copies of the ANNs as a uniformly weighted mixture of Gaussian distributions. The size of the variance, or equivalently the standard deviation, represents the estimated predictive uncertainty.


\section{\label{sec:NumericalExperiment}Numerical Experiment}
\subsection{\label{subsec:EmulationOfNetworkCalc}Emulating Final Abundances with ANNs}
The basic idea of this work is to emulate the variation in the calculated final $r$-process abundance pattern when varying the nuclear physics inputs of the nuclei of interest. The final abundances as a function of the mass number $A$ are calculated by performing nuclear reaction network calculations. As discussed in Section~\ref{subsec:ReactionNetwork}, performing a nuclear reaction network calculation amounts to solving an initial value problem of a system of ODEs. Therefore, in our work, emulating the abundance calculations means modeling the function $f(\cdot)$ that takes some nuclear physics quantities as inputs and maps the initial abundance pattern a function of the mass number $A$, $\bm{Y_A}(t=t_0)$, to the final abundance pattern $\bm{Y_A}(t=t_f)$, where $t_f$ is sufficiently larger than the timescale of the $r$-process nucleosynthesis, e.g. $t_f = 1\;\mathrm{Gyr}$. 

\begin{figure}
    \centering
    \includegraphics[width=\textwidth]{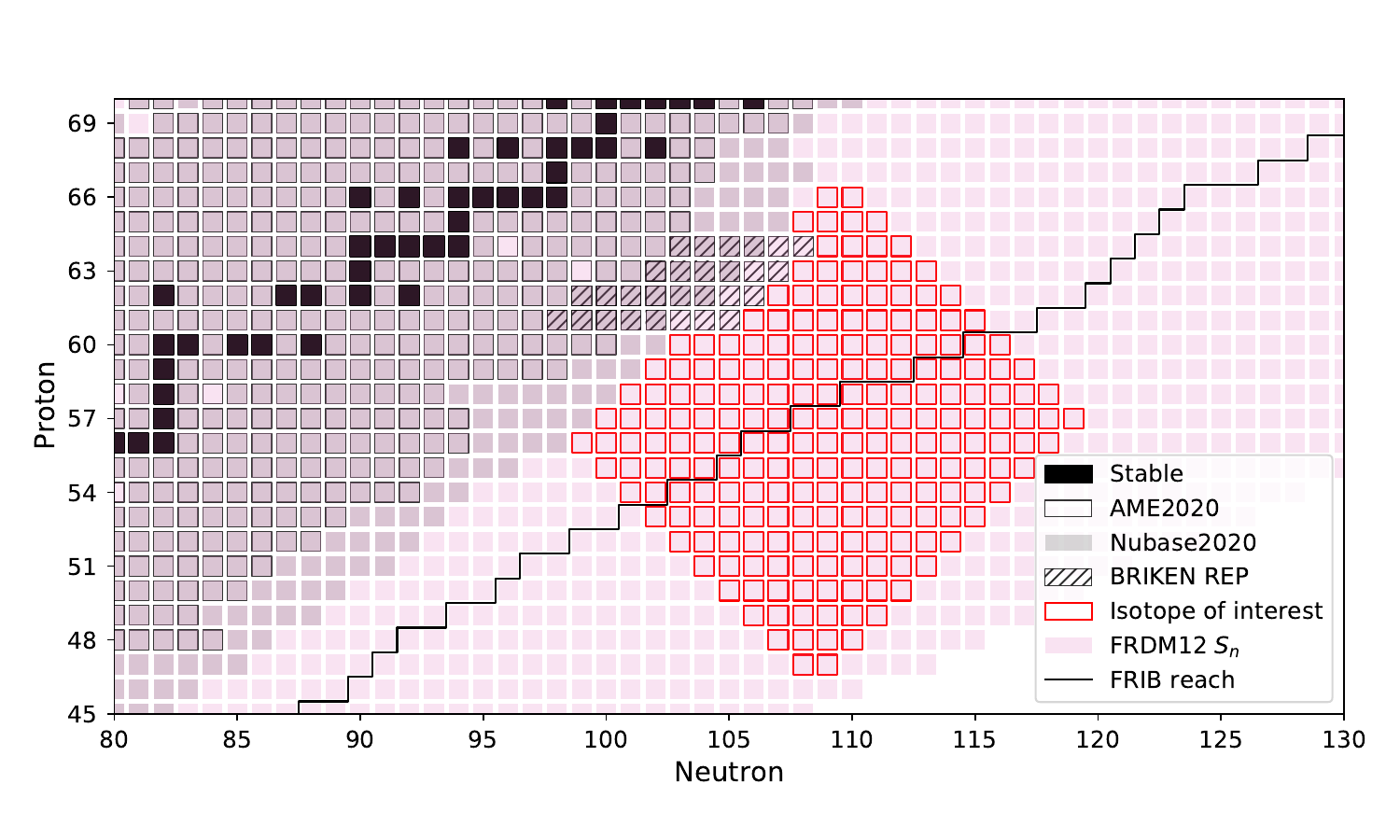}
    \caption[Nuclei included as input variables in the construction of the emulator.]{The rare-earth region of the chart of nuclides. The red squares in the figure shows the nuclei included as input variables in the construction of the emulator. For comparison, the nuclei included in the experimental databases are shown too: AME2020~\cite{AME2020} and Nubase2020~\cite{Nubase2020}. The nuclei measured in our previous work \cite{Kiss_BRIKEN_2022} are shown as well with the label ``BRIKEN REP''. The solid black line represents the estimated reach of radioactive beam production at the Facility for Rare Isotope Beams (FRIB) before the upgrade to FRIB400}.
    \label{fig:Emulator_input_region}
\end{figure}

In this work, we vary one-neutron separation energies ($S_n$) and the $\beta$-decay half-lives ($T_{1/2}$) of the 212 nuclei in the rare-earth region shown in Figure~\ref{fig:Emulator_input_region} and focus on their effect on the rare-earth peak (Section~\ref{subsec:RareEarthPeak}). This model has 424 input variables in total. The neutron separation energies and half-lives are stored as a vector and we denote them as $\bm{S_n}$ and $\bm{T}_{1/2}$, respectively. They are sorted first by the proton number of the nuclei and then the number of neutrons. The function $f(\cdot)$ we aim to model with an ANN is expressed as 
\begin{equation}
    \bm{Y_A}(t=t_f) = f(\bm{S_n},\bm{T}_{1/2}).
\end{equation}
Note that the function $f(\cdot)$ is conditional on the initial abundance pattern $Y_A (t=t_0)$, astrophysical trajectory, other nuclear physics inputs, and all the other inputs of the nuclear reaction network calculations. 

\subsection{\label{subsec:Emulator-distribution}Distributions of theoretical nuclear physics inputs}
For the training of an ANN emulator, we first need to define how the input variables, namely each $S_n$ and $T_{1/2}$, are distributed. If the distributions of the variables of interest are known from uncertainty quantification of theoretical models or from experimental uncertainties, they may be used. However, in general, theoretical nuclear data typically used in nucleosynthesis studies do not have uncertainty estimates, including the FRDM2012 mass model \cite{FRDM2012} and the $\beta$-decay half-lives from FRDM+QRPA \cite{Moller_qrpa_2019} used in this work. As theoretical uncertainties of $S_n$ and $T_{1/2}$, we employ the distributions introduced in Ref.~\cite{Mumpower2016}. The size of the uncertainties for the $S_n$ values are assumed to be $\pm 0.5$~MeV, uniformly distributed around the FRDM values. For the $\beta$-decay half-lives ($T_{1/2}$), we assume that the decay rates ($ \lambda = \ln(2)/T_{1/2}$) are distributed according to log-normal distributions:
\begin{eqnarray}
    p(\lambda) = \frac{1}{\lambda\sqrt{2\pi}\sigma}\exp\left[ -\frac{(\mu-\ln{(\lambda)})^2}{2\sigma^2} \right] \label{eq:lognormal},
\end{eqnarray}
where $\mu$ is the theoretical rate from the FRDM+QRPA prediction and $\sigma^2$ is the variance of the underlying normal distribution, which is set to $\sigma = \ln(2)$ to allow for a factor of 10 in the decay rate variation. 

Practically, however, if samples drawn from the distributions of the variables are directly used for training of the ANNs, it is likely that the tails of the distributions do not have a sufficient number of samples. This would be especially the case for the log-normal distributions introduced above. Therefore, we replace the log-normal distributions with log-uniform distributions that cover the $\pm 3.5\sigma$ intervals of the underlying normal distributions described by Eq.~\ref{eq:lognormal}:
\begin{equation}
    {p\left(\ln(\lambda)\right)} = \cases{
        \frac{1}{7\sigma}  \quad & for $\ln(\lambda)\in [\mu-3.5\sigma,\; \mu+3.5\sigma]$,\\ 
        0 \quad & otherwise.
        \label{eq:uniform_beta}
    }
\end{equation}
To ensure that the samples evenly cover the entire variable space, we employ Sobol sequences \cite{Sobol1976, Saltelli2010}. Sobol sequences are a type of quasi-random sequence designed to fill multidimensional variable spaces as uniformly as possible. Although termed ``quasi-random'', the generated points depend on previously sampled points and fill the gaps between them. For more details, see Ref.~\cite{Saltelli2010} and references therein.


\subsection{\label{subsec:preprocessing}Data pre-processing and training of ANNs}
In order for ANNs to achieve optimal performance, it is necessary to pre-process the input data \cite{Sola_preprocessing_1997}. The main strategy for input data pre-processing in this work is standardization, which makes the input samples distribute with zero means and standard deviations of one. For the $\beta$-decay rates $\lambda$, standardization is performed on a logarithmic scale. Standardization of one-neutron separation energies ($S_n$) is performed on a linear scale. The standardization for each input variable is:
\begin{eqnarray}
    \bar{p}(\lambda) &= \frac{\ln(p(\lambda))-\ln(\lambda^{\mathrm{th}})}{\sigma_{\ln\lambda}^{\mathrm{sample}}},\\
    \bar{p}(S_n) &= \frac{p(S_n) - {S_n}^{\mathrm{th}}}{\sigma_{S_n}^{\mathrm{sample}}},
\end{eqnarray}
where $\lambda^{\mathrm{th}}$ and ${S_n}^{\mathrm{th}}$ denote the theoretical predictions of the FRDM+QRPA model~\cite{Moller_qrpa_2019} and the FRDM2012 mass model~\cite{FRDM2012}, respectively, $\sigma_{\lambda}^{\mathrm{sample}}$ and $\sigma_{S_n}^{\mathrm{sample}}$ are the standard deviations of the sample distributions of $\lambda$ and $S_n$, respectively.

Training has been performed using a type of stochastic gradient descent method called \textsc{AMSGrad}~\cite{AMSGrad2019}, which is a variant of one of the most commonly used methods called \textsc{Adam}~\cite{Adam2014}. Training of our ANNs has been done with 300k samples, of which 280k have been used to optimize the weights in the ANN, and the remaining 20k samples have been used for validation to check the performance of the ANN for unseen input data. 10k samples have been additionally generated after the training is complete, to be used as a test data set for performance evaluation.

We consider two astrophysical trajectories, which provide the temporal evolution of density and temperature: one from cold neutron-rich dynamical ejecta of a binary neutron star merger and the other from a hot wind. These trajectories are identical to those used in Ref.\cite{Kiss_BRIKEN_2022}. The neutron star merger trajectory is from Ref.~\cite{Vassh2019} based on the simulations by Refs.~\cite{Rosswog2013} and \cite{Piran2013}. It takes into account the effect of self-heating based on the FRDM2012 mass model \cite{FRDM2012}. In this trajectory, the temperature decreases rapidly and photodissociation is suppressed (Eq.\ref{eq:detailedbalance}), bringing the path of the nucleosynthesis to the neutron dripline where $S_n\sim0$. After the freeze-out, the $\beta$-decay becomes dominant while neutron capture and photodissociation are further suppressed. The hot wind trajectory corresponds to a hot $r$-process condition with low entropy of $S$=30~$k_B$, an initial electron fraction of $Y_e=0.20$, and an expansion timescale of 70~ms based on Ref.~\cite{Meyer2002}, which is discussed in more detail in Ref.~\cite{Mumpower2016}. In this trajectory, temperature stays sufficiently high to establish the $(n, \gamma) \leftrightarrows (\gamma, n)$ equilibrium. Therefore, the abundance pattern is expected to be more affected by the variation of the $S_n$ values compared to the cold neutron star merger trajectory.

\section{\label{sec:Results}Results and Discussion}
\subsection{\label{subsec:Arch}Optimized Emulator Architecture}
Since no previous literature has been found on emulating nuclear reaction network calculations with ANNs to the best of our knowledge, we employ a systematic and automated way to explore optimal ANN architectures to establish the starting point for the architecture optimization. For this purpose, we use a method called neural architecture search (NAS), implemented in a library called \textsc{AutoKeras} \cite{AutoKeras}. \textsc{AutoKeras} systematically varies the architecture and automatically records the best performing model. The architecture search is guided by Bayesian optimization, which allows for an efficient exploration of the neural network architecture. 

Based on the best performing architecture found by the NAS, we further tuned the architecture manually, mostly by changing the number of layers, the number of filters in the convolutional layers, and the number of units in the fully connected layers.

The results of the neural network architecture optimization done by the NAS and manually are summarized in Table~\ref{tab:architecture}. Our best performing architecture consists of convolutional layers followed by fully connected (dense) layers. In total, including the ``Flatten'' layer, which converts the stacked 2D data into a single vector, there are 7 layers. We have found that the use of convolutional layers is essential for achieving satisfactory performance. The advantage of using convolutional layers is that they can take into account the correlation between the properties of neighboring nuclei on the chart of nuclides. We have experimented with different unit sizes for the sixth fully connected layer, and found that using 1024 units resulted in the optimal performance. As an activation function, ReLU has been used for all layers except the final layer. For the final layer, a linear activation was used to allow unbounded output values. Since the output of this layer is simply a (weighted) linear combination of the output of the previous layer, it is called a ``linear'' layer. Note that the architecture does not have any physical interpretation. The performance of the ANN model is evaluated in detail in the following sections.

\begin{table}[b]
\caption{\label{tab:architecture}%
Architecture of the neural network optimized by neural architecture search then by hand.
}
\begin{tabular}{cccccc}
Layer No.&
Layer type&
Activation&
Kernel size&
No. of filters&
No. of units\\
1 & Convolutional   & ReLU  & (3,3) & 128   & ---\\
2 & Convolutional   & ReLU  & (3,3) & 128   & ---\\
3 & Convolutional   & ReLU  & (3,3) & 128   & ---\\
4 & Convolutional   & ReLU  & (3,3) & 128   & ---\\
5 & Flatten         & ---   & ---   & ---   & ---\\
6 & Fully connected & ReLU  & ---   & ---   & 1024\\
7 & Fully connected & Linear& ---   & ---   & 31\\
\end{tabular}
\end{table}

\subsection{\label{subsec:Performance}Performance}
\begin{figure}[hb]
\includegraphics[width=\textwidth]{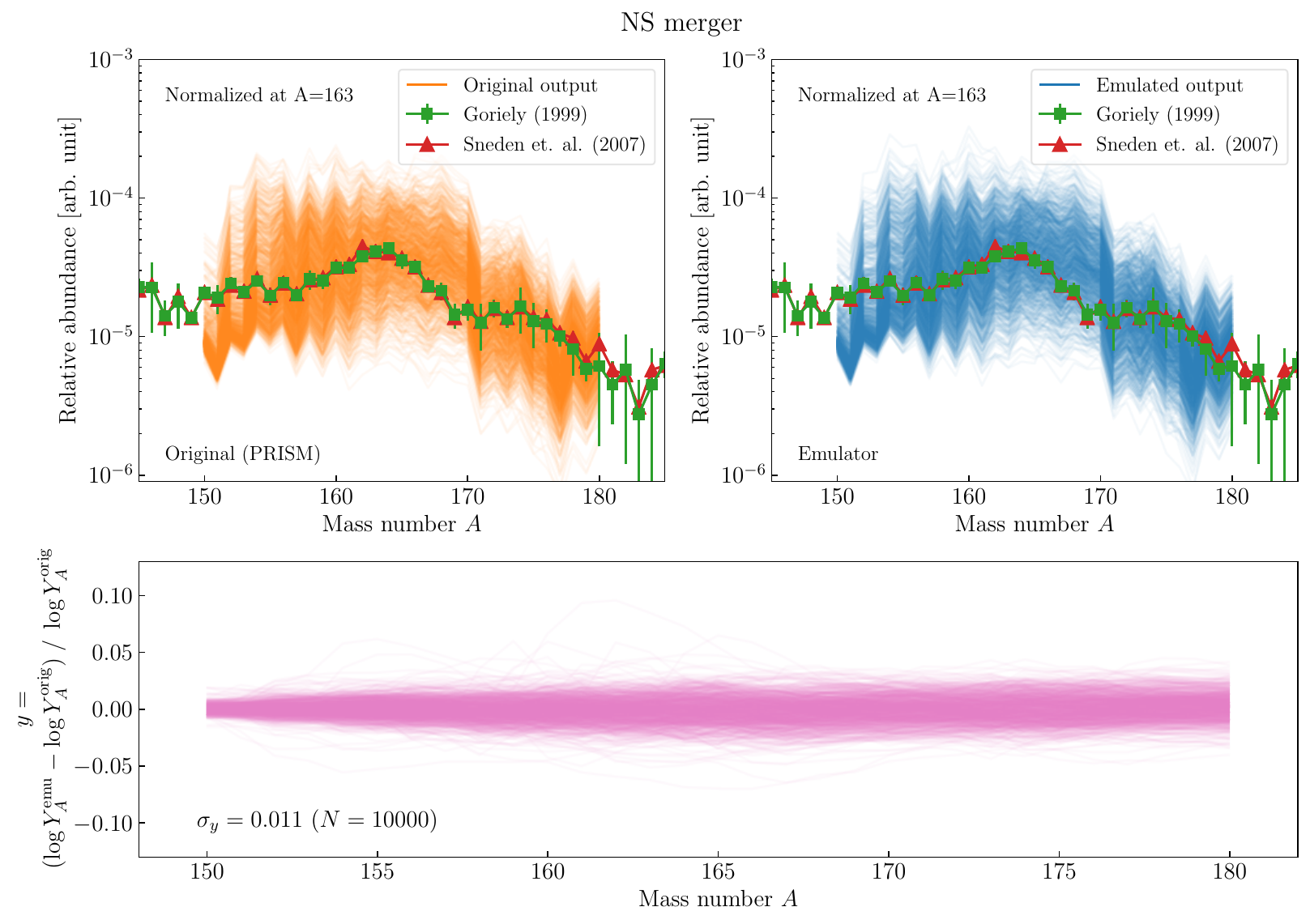}
\caption[Comparison of the results with the unseen test data set between the original nuclear network calculation by \textsc{PRISM} \cite{Mumpower_betadelayedfission_2018} and our ANN emulator for the neutron star merger scenario, focusing on the rare-earth peak (REP) region $A=150$--$180$.]{\label{fig:comparison-new_nsns}Comparison of the results with the unseen test data set between the original nuclear network calculation by \textsc{PRISM} \cite{Mumpower_betadelayedfission_2018} (top left panel) and our ANN emulator (top right panel) for the neutron star merger scenario, focusing on the rare-earth peak (REP) region $A=150$--$180$. The bottom panel shows the relative deviations of the output of the emulator from the original (\textsc{PRISM}) calculations. $\sigma_y$ is their standard deviation, calculated using 10k samples. The plot only shows random 1k samples to avoid overcrowding the plot. The solar abundance patterns are from Refs.~\cite{Goriely1999_solar} and \cite{Sneden2008_solar}, and they are scaled to match the average of the \textsc{PRISM} calculations at $A=163$, which is the local abundance maximum in this mass region.}
\end{figure}

\begin{figure}[hb]
\includegraphics[width=\textwidth]{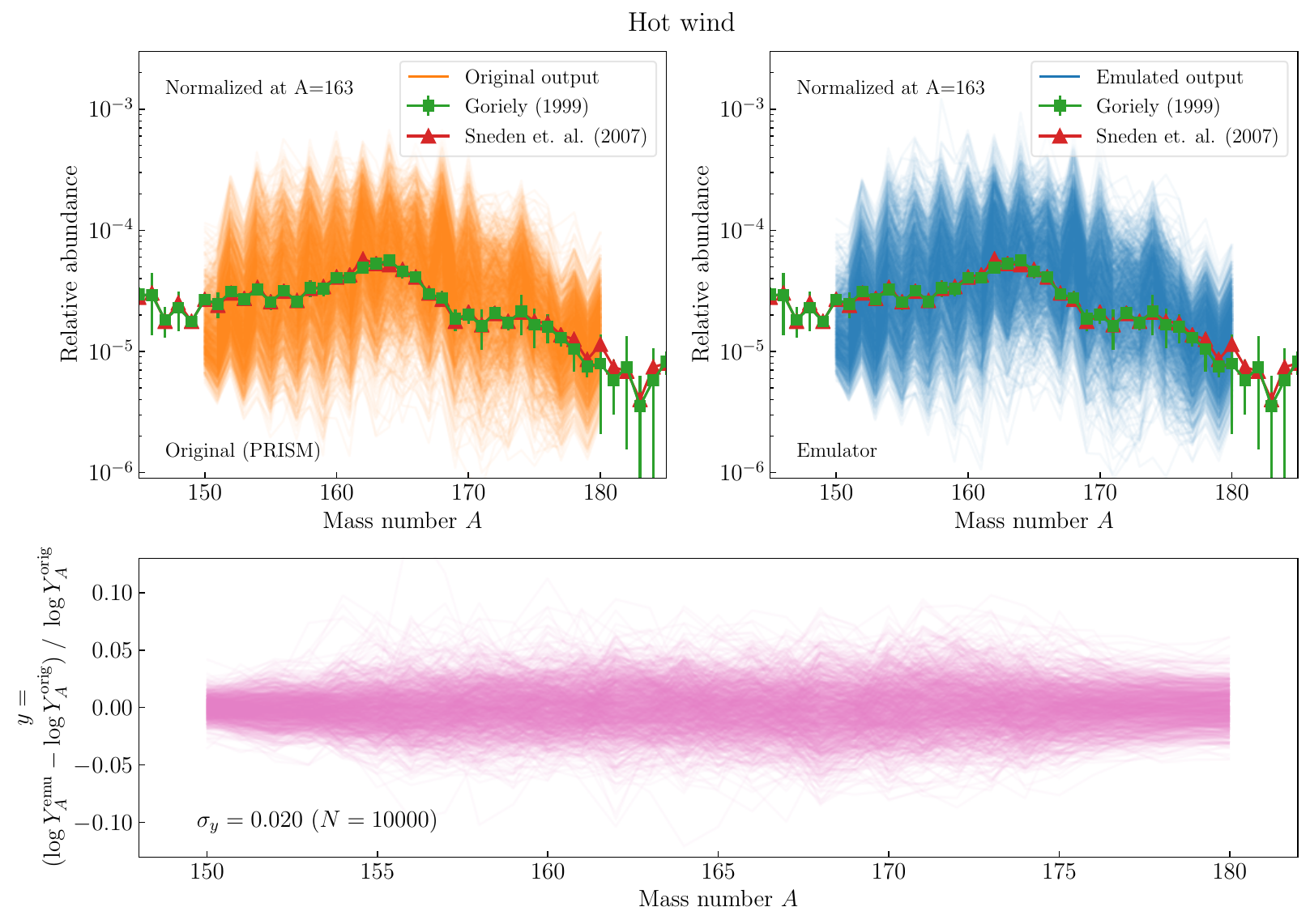}
\caption[Same figure as Figure~\ref{fig:comparison-new_nsns}, but for the hot wind scenario.]{\label{fig:comparison-wind}Same figure as Figure~\ref{fig:comparison-new_nsns}, but for the hot wind scenario.}
\end{figure}

Figures~\ref{fig:comparison-new_nsns} and \ref{fig:comparison-wind} show comparisons between the output of the original nuclear reaction network calculations (\textsc{PRISM}) \cite{Mumpower_betadelayedfission_2018} and the output of the ANN emulator, using the test data set consisting of 10k samples for the two astrophysical scenarios employed. Note that the figures only show randomly selected 1k samples to avoid overcrowding the plot. Comparing the top two panels of the figures indicates that the ANN emulator captures the general trends of the calculated abundances very well. 
The bottom panel of the figures shows the deviations of the output of the emulator ($\log{Y_A^{\mathrm{emu}}}$) from the original (\textsc{PRISM}) calculations ($\log{{Y_A^{\mathrm{orig}}}}$), relative to the original calculations, defined as
\begin{equation}
    y \equiv \frac{\log{Y_A^{\mathrm{emu}}}-\log{{Y_A^{\mathrm{orig}}}}}{\log{{Y_A^{\mathrm{orig}}}}}.
\end{equation}
The $\sigma_y$ shown in the bottom panel is the standard deviation of $y$, calculated using the entire 10k test samples. For the neutron star (NS) merger scenario, the standard deviation of the value $y$ is $\sigma_y = 0.011$ (1.1\%). For the hot wind scenario, it is $\sigma=0.02$ (2\%). The larger variation of the abundances in the hot wind scenario is most likely because the $(n,\;\gamma)\leftrightarrows (\gamma, \; n)$ equilibrium is established, which is affected by the neutron separation energies. In the NS merger scenario, due to its extremely neutron-rich and cold condition, the path of the $r$-process nucleosynthesis is pushed all the way to the neutron dripline where $S_n \sim 0$. Therefore, the photodissociation rates are highly suppressed due to Eq.~\ref{eq:detailedbalance} and the final abundance pattern is less affected by the variation of the $S_n$ values of the nuclei that are far from the dripline.

The main advantage of using emulators is their speed. While a nuclear reaction network calculation is not an extremely computationally expensive calculation, a single run of \textsc{PRISM} for the neutron star merger scenario takes roughly 400 seconds on an Intel Xeon CPU E5-2683 v4, available on the compute cluster \textsc{Graham} of the Digital Research Alliance of Canada. Multiple calculations can be run independently in parallel, but each run requires a compute core and about 4~GB of memory. On the other hand, obtaining a single abundance pattern from our emulator only takes about 0.02~seconds on average per evaluation over 10k evaluations, using a NVIDIA Tesla P100 GPU, also available on \textsc{Graham}. For a single abundance calculation, this is a speed-up by a factor of 20,000. Furthermore, returning outputs for multiple input samples is also efficient---it takes about 6~second to predict 10k abundance patterns for the test data set.

\subsection{\label{subsec:UQ}Uncertainty Quantification}
\begin{figure}[hb]
\centering
\includegraphics[width=0.6\textwidth]{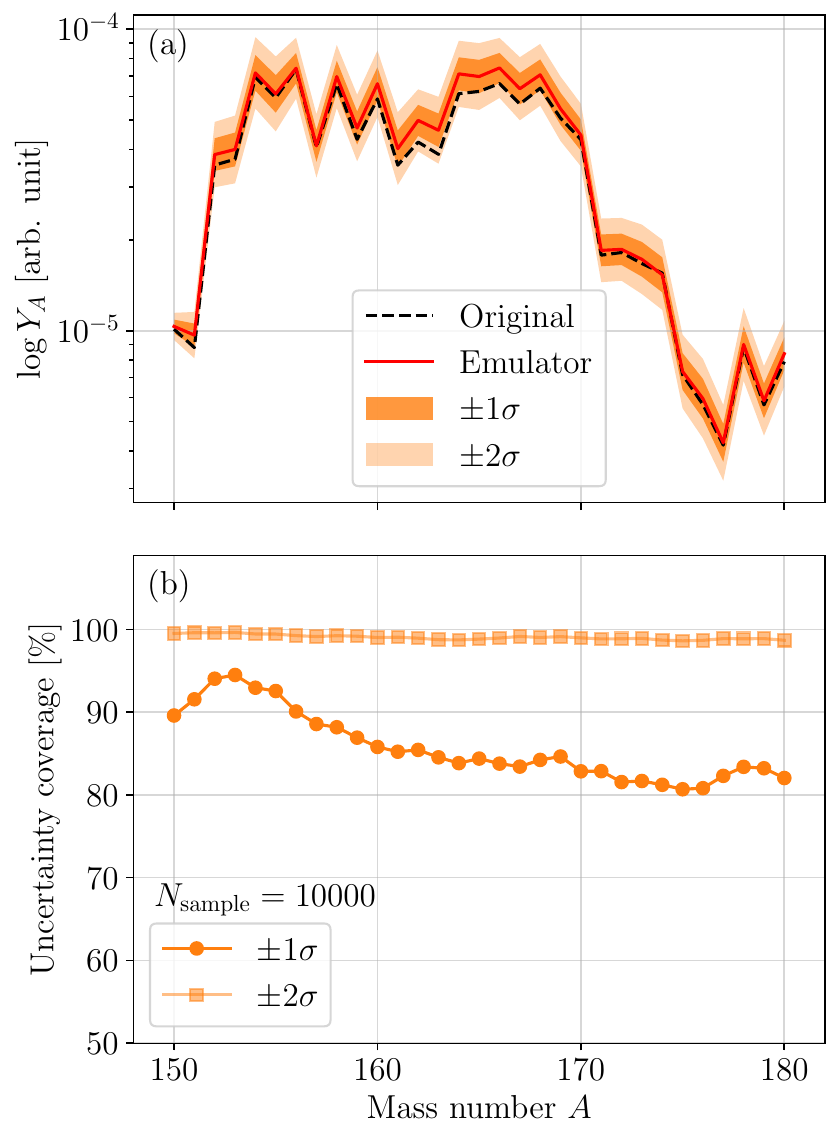}
\caption[Panel (a): the emulated abundance pattern and the estimated $\pm 1 \sigma$ and $\pm 2 \sigma$ uncertainty bands of one of the test samples, compared to the original (\textsc{PRISM}) calculation. Panel (b): how many of the original calculations of the 10k test samples are covered by the $\pm 1 \sigma$ and $\pm 2 \sigma$ uncertainty bands, respectively.]{\label{fig:uncertainty}Panel (a) shows an example of the emulated abundance pattern and the estimated $\pm 1 \sigma$ and $\pm 2 \sigma$ uncertainty bands of one of the test samples, compared to the original (\textsc{PRISM}) calculation. Panel (b) shows how many of the original calculations of the 10k test samples are covered by the $\pm 1 \sigma$ and $\pm 2 \sigma$ uncertainty bands, respectively.}
\end{figure}

Uncertainty quantification of ANN predictions has been performed using deep ensembles \cite{DeepEnsemble}. The top panel of Figure~\ref{fig:uncertainty} shows an example of the uncertainty quantification of an ANN prediction for the abundance pattern in the REP region (mass number $150\leq A \leq 180$) drawn from the test dataset, compared to the original abundance pattern calculated with \textsc{PRISM}. It shows that the size of the uncertainty band is small enough to resolve the details of the abundance pattern. The bottom panel of the same figure shows how many of the 10k test samples of the original calculations are covered by the $\pm 1 \sigma$ and $\pm 2 \sigma$ uncertainty bands. Since our uncertainty is assumed to follow a Gaussian distribution, roughly 68~\% and 95~\% of the data points are expected to be covered by the $\pm 1 \sigma$ and $\pm 2 \sigma$ uncertainty bands, respectively. In our numerical experiment, it can be seen from the figure that about 80-94~\% and 98-99~\% of the original calculations are covered by the $\pm 1 \sigma$ and $\pm 2 \sigma$ uncertainty bands, respectively. This means that our uncertainty bands are somewhat under-confident (the size of uncertainty is overestimated); nevertheless, this simple method can provide meaningful estimates of prediction uncertainty.

\section{\label{sec:example}An Example: Propagation of Nuclear Physics Input Uncertainty} Nuclear reaction network calculation emulators can speed up various computationally intensive tasks, such as inverse problems, uncertainty propagation, and sensitivity analyses. In this section, propagation of the nuclear physics input uncertainty is demonstrated as an example of such tasks. 

\begin{figure}[hb]
\centering
\includegraphics[width=0.6\textwidth]{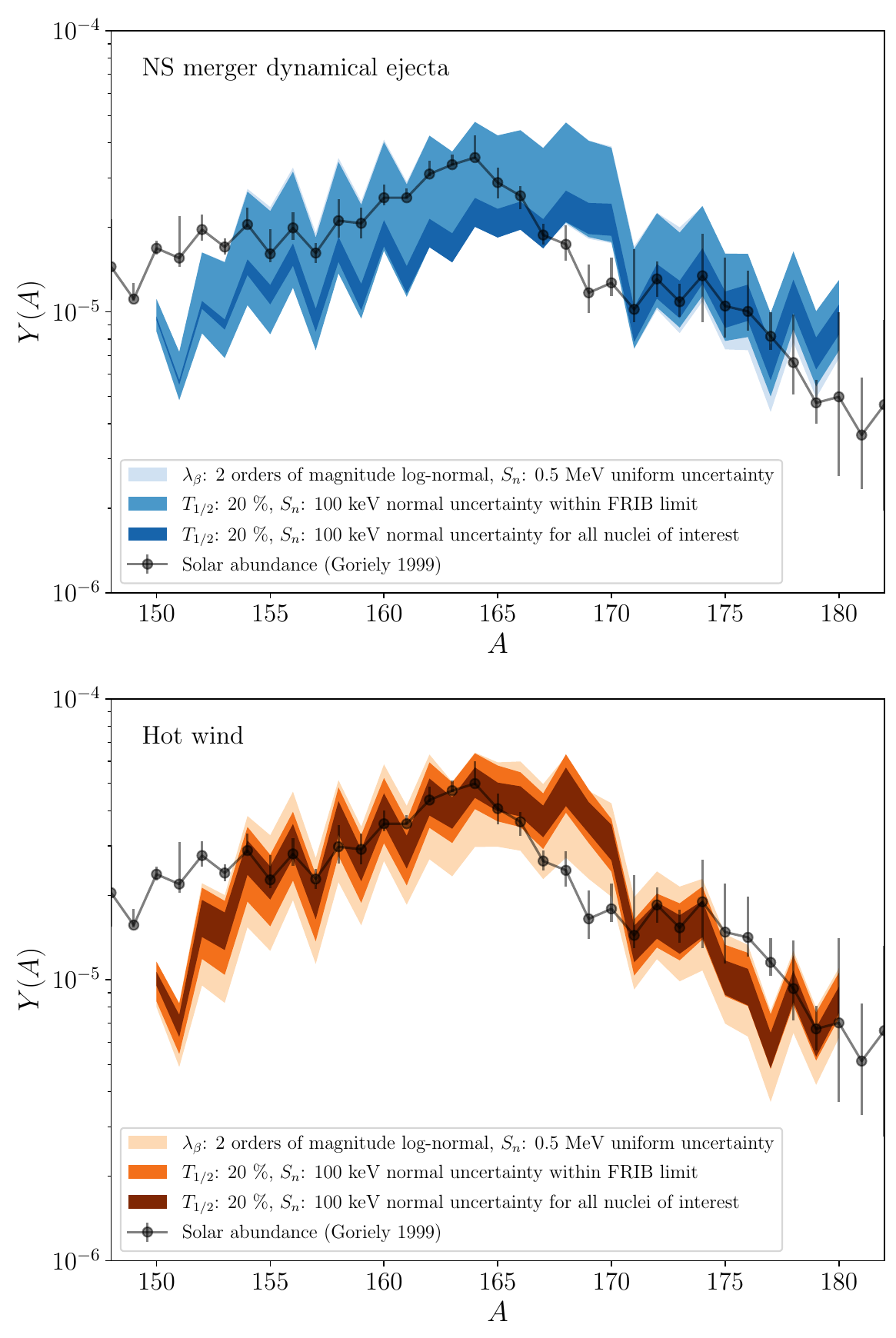}
\caption{\label{fig:propagated_uncertainty}
Uncertainty bands of abundance patterns propagated from the uncertainties of one-neutron separation energies $S_n$ and $\beta^{-}$-decay half-lives $T_{1/2}$ (or equivalently decay rates $\lambda$) using emulators. The top panel shows the results for the neutron star dynamical ejecta condition and the bottom panel shows the hot wind condition. In the top panel, the band with the lightest blue color is not visible except for $172\leq A \leq 180$, since it is mostly overlapping with the band with the medium-light blue color. The solar abundance pattern and its uncertainty \cite{Goriely1999_solar} is also shown for comparison.
}
\end{figure}

The nuclei of interest are the same as the input of the emulator, as shown in Fig.~\ref{fig:Emulator_input_region}, and their one-neutron separation energies ($S_n$) and $\beta^{-}$-decay half-lives ($T_{1/2}$) are the inputs. We consider the following three hypothetical cases for the uncertainties of $S_n$ and $T_{1/2}$:
\begin{enumerate}
    \item For all nuclei of interest, $S_n$ is a uniform distribution around the FRDM2012 masses with a range of $\pm 0.5$~MeV, and the $\beta^{-}$-decay rate $\lambda = \ln{(2)}/T_{1/2}$ is a log-normal distribution (Eq.~\ref{eq:lognormal}) with $\mu$ being the FRDM+QRPA prediction and $\sigma = \ln{(2)}$. This is the same as the test data.
    \item For the nuclei within the FRIB reach shown in Fig.~\ref{fig:Emulator_input_region}, assume that $S_n$ is a normal distribution with a standard deviation of $\pm 100$~keV around the FRDM2012 prediction and that $T_{1/2}$ is a normal distribution with the mean being the FRDM+QRPA prediction and the standard deviation being 20\% of the mean. For other nuclei, the distributions are the same as in (i).
    \item For all nuclei of interest, $S_n$ is a normal distribution with a standard deviation of $\pm 100$~keV around the FRDM2012 prediction and $T_{1/2}$ is a normal distribution with the mean being the FRDM+QRPA prediction and the standard deviation being 20\% of the mean.
\end{enumerate}
In cases (ii) and (iii), the normal distributions represent the hypothetical experimental uncertainty, the predictions by FRDM2012 and FRDM+QRPA being the hypothetical nominal experimental values. Since the distributions in cases (ii) and (iii) are narrower than those in case (i), which are identical to the test data, the performance of the emulator demonstrated in Section \ref{subsec:Performance} applies. 10k samples are generated from these distributions (each sample consists of the $S_n$ and $T_{1/2}$ of the 212 nuclei), and the nucleosynthesis yields in the rare-earth region have been predicted with the emulator. The propagated uncertainty bands in the three cases are shown in Fig.~\ref{fig:propagated_uncertainty}, for the neutron star merger dynamical ejecta condition (top) and the hot wind condition (bottom).

For both astrophysical conditions, it can be seen that the nuclear physics input uncertainty described in case (i) results in significant uncertainties in the abundance pattern in the rare-earth region compared to the observed solar abundance pattern. When the uncertainty is reduced for the nuclei within the FRIB reach, there is a significant reduction in the abundance pattern uncertainty for the hot wind condition, but the reduction is minimal in the dynamical ejecta condition. This is because, in the hot wind condition, ($\gamma, n$) photodissociation is enhanced due to the higher temperature and the $r$-process path is determined by the $(n, \gamma)\leftrightarrows (\gamma, n)$ equilibrium, which depends sensitively on the $S_n$ values (Saha equation). In the cold and neutron-rich dynamical ejecta condition, photodissociation is highly suppressed and the path lies very close to the one-neutron drip line. This is why the reduction in uncertainty is only visible when the nuclear physics uncertainty is also reduced for more neutron-rich nuclei, as in case (iii).

The amount of time it took the emulator to predict the 30k abundance patterns (10k each for the three cases in one astrophysical condition) was about 19~seconds on a Nvidia GPU Tesla P100. A single nuclear reaction network calculation with the same input takes 1-10 minutes; therefore, calculating 30k abundance patterns would take 500-5000 core hours. Of course, a large number of samples are required to train the emulator; however, once the training is complete, one can quickly perform computationally intensive tasks as demonstrated in this example.

\section{\label{sec:Conclusions}Conclusions}
In this work, we have shown that it is possible to emulate the calculation of final abundance pattern with traditional ANNs consisting of convolutional layers followed by fully connected layers. Emulators have been constructed for two astrophysical scenarios: neutron star mergers and the hot wind. The performance of the emulator has been demonstrated focusing on the rare-earth peak region ($150\leq A \leq 180$), by treating the $\beta$-decay rates and the one-neutron separation energies of 212 isotopes as input variables for our ANN (in total 424 input variables). For both astrophysical trajectories, the ANNs can approximate the original calculations by the nuclear reaction network calculation code \textsc{PRISM} with less than 5~\% deviation.  

We demonstrated estimation of the predictive uncertainty of the ANN using deep ensembles, and the quality of the uncertainty estimation has been evaluated. The method provides conservative but meaningful uncertainty bands. 

Dramatic speed-up of a single $r$-process abundance calculation, roughly by a factor of 20,000, has been achieved. The emulator can also predict a large number of abundance patterns at once in a short amount of time. This implies that large-scale statistical tasks that require performing nuclear reaction network calculations repeatedly, such as uncertainty quantification, inverse problems, and variance-based sensitivity analyses, can be performed significantly faster, as illustrated by the example of the nuclear physics input uncertainty propagation.

Although this work has shown that it is possible to create an emulator for abundance calculations by focusing on the properties of rare-earth nuclei and their final abundances, further development is required to create an emulator that can handle the entire chart of nuclides. This is a challenging task due to the high-dimensional input space; therefore, it will most likely require a clever dimension reduction method as well as efficient learning algorithms. For a full emulation of nuclear reaction network calculations, the time dependence of nuclear abundances must be emulated. The ability to take astrophysical conditions (temporal evolution of temperature and density) as variable input would also be desirable.


\section{Acknowledgments}
This research was enabled in part by support provided by the Digital Research Alliance of Canada (alliancecan.ca) and the BC DRI Group.
Y.S. and I.D. acknowledge funding from the Canadian Natural Sciences and Engineering Research Council (NSERC), NSERC Discovery Grant No.~SAPIN-2019-00030, No.~SAPPJ-2017-00026, and the NSERC CREATE Program IsoSiM (Isotopes for Science and Medicine).
R.K. is supported by the U.S. Department of Energy, Office of Science, Office of Nuclear Physics under Contract No. DE-AC02-05CH11231.
M.R.M. is supported by the U.S. Department of Energy
through the Los Alamos National Laboratory (LANL). LANL
is operated by Triad National Security LLC for the National
Nuclear Security Administration of the U.S. Department of
Energy (Contract No. 89233218CNA000001).
Y.S. and R.S. acknowledges support from the U.S. National Science Foundation under grant number 21-16686 (NP3M).
R.S. acknowledges support from the U.S. Department of Energy contract DE-FG02-95-ER40934.

\section*{References}
\bibliography{main}

\end{document}